\documentclass[12pt,preprint2]{aastex}

\input{psfig.sty} 

\begin{document} 

\title{Is the Lack of Pulsations in Low Mass X-Ray Binaries due to
Comptonizing Coronae?} 

\author{
Ersin {G\"o\u{g}\"u\c{s}}\altaffilmark{1}, 
M. Ali Alpar\altaffilmark{1},
Marat Gilfanov\altaffilmark{2,3}
}

\altaffiltext{1}{Sabanc{\i} University, Faculty of Engineering \& Natural
Sciences, Orhanl{\i}$-$Tuzla 34956 {\.I}stanbul, Turkey}  
\altaffiltext{2}{Max-Planck-Institute f{\"u}r Astrophysik, 
Karl-Schwarzschild-Str. 1, 85740 Garching bei M{\"u}nchen, Germany}
\altaffiltext{3}{Space Research Institute, Russian Academy of Sciences, 
Profsoyuznaya 84/32, 117997 Moscow, Russia}


\begin{abstract} 

The spin periods of the neutron stars in most Low Mass X-ray Binary
(LMXB) systems still remain undetected.
One of the models to explain the absence of coherent pulsations has been 
the suppression of the beamed signal by Compton
scattering of X-ray photons by electrons in a surrounding corona. 
We point out that simultaneously with wiping out the pulsation signal,
such a corona will upscatter (pulsating or not) X-ray emission
originating at and/or near the surface of the neutron star leading to
appearance of a hard tail of Comptonized radiation in the source
spectrum. We analyze the hard X-ray spectra of a selected set of LMXBs 
and demonstrate that the optical depth of the corona is not likely to be 
large enough to cause the pulsations to disappear.

\end{abstract} 

\keywords{accretion, neutron star physics -- X-rays: Binaries --
stars: individual (GX 9+1), (GX 9+9), (Sco X-1)}

\section{Introduction} 

Low mass X-ray binaries (LMXBs) are binary systems that contain a
neutron star or a black hole as the primary object, and a low-mass star 
(typically M $<$ 2 M$_{\rm Sun}$) as the mass-donating companion. Mass 
transfer from the companion takes place via Roche lobe overflow. 

In the case of neutron star LMXBs, the neutron star spin period would be 
expected to be observable if the magnetic fields of the star can channel 
the accretion to yield beamed X-ray emission, and if the beamed signal 
can escape the system. The fact that almost none of the LMXBs exhibit 
pulsations corresponding to the neutron star spin period has remained a 
puzzle since the discovery of neutron star LMXBs. The proposal that the 
millisecond radio pulsars are evolutionary 
descendants of the low mass X-ray binaries (Alpar et al. 1982; Radhakrishnan 
\& Srinivasan, 1983) highlighted the issue and led to a search for millisecond 
X-ray pulsars. Millisecond spin periods were also indicated in the initial 
beat frequency model for horizontal branch quasi periodic luminosity
oscillations (Alpar \& Shaham 1985). 
More recently, kilohertz QPO branches also gave an 
indication of millisecond rotation 
periods through a beat frequency model interpretation of the difference between 
two "kilohertz" frequency bands. The burst oscillations observed from these 
sources seem to have about the difference frequency or half the difference 
frequency. While a consistent interpretation of kilohertz, burst and horizontal 
branch oscillations is not available yet, there are enough correlations and 
systematics (van der Klis 2000) at high frequency bands that, assuming the 
accretion flow 
and the neutron star are not far from rotational equilibrium, all these high 
frequency QPO observations also point at rapid neutron star spin, with periods 
in the millisecond range.

Millisecond X-ray pulsars were discovered relatively recently (Wijnands \&
van der Klis 1998), 
confirming the evolutionary hypothesis. These millisecond X-ray pulsars 
(or, indeed , pulsars of any coherent spin period) remain a minority 
among the LMXBs. 
Almost 90\% of LMXBs do not display coherent pulsations in their persistent 
phase. So the question remains as to why these few among the LMXBs display 
their spin period. 

The millisecond radio pulsars are believed to have spun up to these exremely 
short spin periods by accretion. To have millisecond equilibrium periods at 
sub-Eddington accretion rates the magnetic field strength of the compact 
object should be relatively weak (B $\sim$ 10$^9$ G). Nevertheless, a magnetic 
field of this magnitude may be able to channel the accretion flow and lead 
to beamed X-ray radiation. The inner radius of the accretion disk, roughly the 
Alfven radius, is expected to be a few stellar radii in most LMXBs. This 
leaves a possibility of a magnetosphere with ample options for anisotropy in 
accretion and beaming for the X-ray radiation. 

The absence of coherent pulsations in the persistent emission of neutron star
LMXBs is usually attributed to several potential causes as laid out already 
in the first papers on the evolution of millisecond radio pulsars 
(Alpar et al. 1982, Alpar and Shaham 1985, Lamb et al 1985):
{\it (i)} The magnetic field is so weak that accreting matter cannot be 
channeled onto the magnetic poles, {\it (ii)} The pulsar's periodic signal 
is attenuated by gravitational lensing (e.g. Meszaros, Riffert \& Berthiaume 
1988), {\it (iii)} Beamed radiation emerging from the neutron star's magnetic 
polar caps is "wiped out" by electron scattering (Brainerd \& Lamb 1987; 
Kylafis \& Klimis 1987; Wang \& Schlickeiser 1987; Bussard et al. 1988).
It is important to note the distinction between isotropic luminosity 
oscillations and modulation in the observed signal due to a rotating beam. 
Although in both cases scatterings will suppress pulsations as observed by a
distant observer, the physics of suppression is different
(e.g. Miller, Lamb, \& Psaltis 1998). In the case of isotropic luminosity 
oscillations, the degree of suppression of pulsations depends on the ratio 
of the light travel time in the scattering media (which depends on its optical 
depth and size) to the pulsation period. In the case of modulations due to 
rotating beamed radiation, the optical depth of $\sim 1$ is sufficient to 
destroy  the beaming and to wipe out pulsations. While the luminosity 
oscillation may be relevant to QPO phenomena in X-ray binaries, it is the 
beaming oscillations, that are responsible for coherent pulsations observed 
from X-ray pulsars. The latter case is critically investigated in the paper.

In the commonly accepted picture of sub-Eddington accretion onto a
neutron star, two parts of the accretion flow are distinguished - the
Keplerian accretion disk and the boundary/spreading layer on the
surface of the  neutron star (Sunyaev \& Shakura 1986). 
The boundary/spreading layer is present in neutron stars with 
magnetic field not strong enough to stop the acretion disk beyond a 
magnetosphere. In the boundary/spreading layer the
accreting matter decelerates from the Keplerian velocity of the inner
boundary of the accretion disk to the rotation velocity of the neutron star 
and settles onto its surface. Comparable fractions of energy are emitted
in the disk and in the boundary layer (Sibgatullin \& Sunyaev 2000). 
Correspondingly, there are two components in the spectra of neutron 
star binaries, associated with these two parts of the
accretion flow. At sufficiently high mass accretion rate $\dot{\rm M}>$ 0.1
$\dot{\rm M}_{\rm Edd}$, corresponding to the high spectral state of atoll and
Z-sources, the boundary layer and accretion disk are both radiating in
the optically thick regime with kT$\sim$1$-$2 keV. It is expected on
theoretical grounds and demonstrated observationally that the
boundary layer component has higher temperature than the accretion
disk (Gilfanov, Revnivtsev \& Molkov 2003).

A corona around the neutron star, responsible for wiping out the pulsating 
signal will upscatter the relatively soft emission from the neutron star 
surface (pulsating or not). Depending on the geometry, some fraction of 
the accretion disk emission will also be upscattered.
This will lead to appearance of the hard tail of Comptonized emission
in the spectrum. Transient hard tails are indeed observed in the
spectra of neutron star LMXBs and are sometimes attributed to the
Comptonization (see e.g., Di Salvo et al. [2000]). However, these hard 
tails are not always present.

Hard X-ray upper limits (or detected flux) can be used to constrain
parameters of the putative corona around the neutron stars. Indeed, if the 
spectrum and intensity of the neutron star surface emission are given, the 
flux in a high energy band (e.g. 30$-$60 keV) will depend uniquely on the 
temperature and optical depth of the corona. The critical parameter in the 
"wiping out" scenario is, of course, the optical depth. One can consider a 
range of plausible temperatures and constrain the optical depth of the corona
as a function of its (unknown) temperature. If, on the other hand, the
hard tail is significantly detected, the position of its  high energy
cut-off can help to determine the temperature and more accurately
constrain the optical depth.

In this study, we perform broadband X-ray spectral investigations of the 
three LMXBs, GX 9+1, GX 9+9 and Sco X-1 to constrain their hard X-ray spectral
characteristics using a Comptonization model. We estimate the spectrum and flux of 
the seed photons for Comptonization (i.e. neutron star surface spectrum) by 
applying a two-temperature black body model to the observed spectrum below
$\sim$20 keV. In this simplified model, the harder blackbody represents
emission of the surface of the neutron star (boundary/spreading layer). 
The softer blackbody is a crude approximation to the emission of the accretion 
disk, ignoring the temperature distribution in the disk and the effect of 
scatterings. The Comptonization model is mainly characterized with the 
usual two parameters, the electron scattering optical thickness $\tau$ and 
the electron temperature kT$_{\rm e}$. Values of $\tau$ $\gtrsim$ 1 would 
destroy the beamed radiation from the surface of the neutron star. 
We can, therefore, check the validity of the electron 
scattering scenario for the absence of coherent pulsations from these systems 
by using the upper limits to hard X$-$ray emission to obtain upper limits
of $\tau$ for a range of electron temperatures and conclude whether 
or not the optical depth of a corona is large enough to smear out the 
beamed radiation from the neutron star.
In the following section we describe the data 
used in this study. We present our spectral investigations in \S 3. 
The results are presented in \S 4, discussion and conclusions in \S 5.

\section{Observations and Data Extraction}

\begin{table}
\begin{center}

\caption{Log of RXTE observations.
\label{tbl:rxte_data}}
\vspace{0.2cm}
\begin{tabular}{lccc}
\hline
\hline
Source           &   ObsID        &     Date     &  Exp (ks) \\
\hline
{$\rm{GX 9+1}$}  & 20064-01-01-00 &  10/02/1997  &    11.8  \\
{$\rm{GX 9+9}$}  & 10072-04-02-00 &  16/10/1996   &    10.1  \\
{$\rm{Sco X-1}$} & 20053-01-01-05 &  23/04/1997     &    15.1  \\
\hline
\hline
\end{tabular}

\end{center}
\end{table}

We used archival Rossi X$-$ray Timing Explorer (RXTE) pointing
observations of GX 9+1, GX 9+9 and Sco X$-$1. These sources are selected
to reflect broad characteristics of LMXBs. GX 9+1 and GX 9+9 are known
as atoll sources and their X-ray intensities are in the range of low (0.1
L$_{\rm Edd}$) to intermediate (0.4 L$_{\rm Edd}$) levels.
Sco X-1 is a Z source which is the brightest X-ray emitting system among 
LMXBs ($\gtrsim$ L$_{\rm Edd}$). In Table 1 we list the 
details of the RXTE pointed observations.

RXTE consists of two main instruments. The Proportional Counter
Array (PCA) is an array of five nearly identical xenon Proportional
Counter Units (PCUs), which are sensitive to photon energies between
2$-$60 keV. Each detector unit has a collecting area of 1300 cm$^2$ and
energy resolution of 18\% at 6 keV (Jahoda et al. 2006).
The High-Energy X-Ray Timing Experiment (HEXTE) consists of two
clusters of NaI/CsI scintillation detectors which are sensitive to
photons of energies between 15 and 250 keV. The energy resolution
of HEXTE detectors is 15\% at 60 keV (Rothschild et al. 1998).

We extracted the PCA spectra using Standard2 data (129 channels accumulated 
every 16 s) collected from all three layers of all five PCUs. 
In selecting data we required the Earth elevation angle with respect to the
spacecraft to be greater than 10$^\circ$ and the time to the nearest 
South Atlantic 
Anomaly passage to be more than 30 minutes. A background spectrum was 
generated using the bright source background models, provided by the PCA 
instrument team and pcabackest, which is an HEASOFT utility. The detector
dead time correction was applied for both the PCA and HEXTE spectra due to very
bright X-ray output of the sources selected. 
We performed X-ray spectral modeling using XSPEC version 11.3.1 (Arnaud 1996). 
We added a 1\% systematic error to the statistical error of each PCA spectral 
channel to account for the detector response uncertainties.

\section{Spectral Analysis} 

We initially modeled the PCA spectrum only (in 3$-$20 keV) to determine the 
spectral characteristics of these sources at low X-ray energies. The hydrogen
column densities (N$_{\rm H}$) were fixed at the interstellar average values
in the direction of each source (Dickey \& Lockman, 1990). 
The interstellar N$_{\rm H}$ values are 
9$\times$10$^{21}$ cm$^{-2}$, 2$\times$10$^{21}$ cm$^{-2}$, and 
1.5$\times$10$^{21}$ cm$^{-2}$ for GX 9+1, GX 9+9, and Sco X-1, respectively. 

The X-ray spectrum of GX 9+1 is well fit with a sum of two blackbody functions.
The blackbody temperatures are 1.96$\pm$0.02 keV and 1.08$\pm$0.02 keV. 
The emission radius of the higher temperature blackbody is 7.8$\pm$0.4 km, 
while the radius of the lower temperature blackbody is 22.8$\pm$0.7 km 
(assuming a distance of 10 kpc). We take the former blackbody component
as the emission originating from nearby the neutron star surface.
Note that these radii may not represent exact emission areas due to effect 
of scattering and orbital inclination.
The unabsorbed 2$-$20 keV flux values of the blackbody components are 
9.14$\times$10$^{-9}$ erg cm$^{-2}$ s$^{-1}$ and 
6.46$\times$10$^{-9}$ erg cm$^{-2}$ s$^{-1}$, respectively.

In case of GX 9+9 spectral fitting, a broad spectral line was required to 
adequately fit the spectrum, as well as two blackbodies. 
The blackbody temperatures are 2.02$\pm$0.06 keV and 0.88$\pm$0.02 keV, and 
the radii of corresponding emitting regions are 3.9$\pm$0.2 km and 
21.6$\pm$0.9 km (assuming a source distance of 10 kpc). 
The centroid energy of the broad line feature was fixed at 6.4 keV (that is, 
the rest frame energy of cold iron). We found a line width of 1.05$\pm$0.03 
keV and the equivalent width of 240$\pm$22 eV.
Similar to the case in GX 9+1, we take the 2.02 keV blackbody as emitting from
on or near the neutron star surface. We estimate the unabsorbed 2$-$20 keV 
flux values as 2.65$\times$10$^{-9}$ erg cm$^{-2}$ s$^{-1}$ and 
1.96$\times$10$^{-9}$ erg cm$^{-2}$ s$^{-1}$ for the harder and softer 
blackbody components, respectively.

As for Sco X-1, the sum of two blackbody components and
a broad line feature yields a suitable fit to the X-ray spectrum. The
blackbody temperatures are 2.27$\pm$0.01 keV and 0.78$\pm$0.01 keV.
The radii of emitting regions are 5.3$\pm$0.3 km and 54.1$\pm$1.5 km
(using the distance estimate of 2.8 kpc [Bradshaw, Fomalont \& Geldzahler 
1999]). The width of the line feature (the centroid fixed at 6.4 keV) is
1.15$\pm$0.1 keV and the equivalent width is 848$\pm$9 eV. The unabsorbed
fluxes in the 2$-$20 keV range are estimated as 
1.13$\times$10$^{-7}$ erg cm$^{-2}$ s$^{-1}$ and 1.05$\times$10$^{-7}$ 
erg cm$^{-2}$ s$^{-1}$ for the harder and softer blackbody components, 
respectively. 

For each source, we then modeled the PCA (3$-$20 keV) and HEXTE (15$-$200 keV) 
spectra simultaneously as follows.
We fixed the spectral parameters of the low energy portion on
their above determined values and added an extra Comptonized
component (COMPPS model in XSPEC, spherical geometry [Poutanen \& 
Svensson 1996]). The spectrum and normalization of the seed photons for the
Comptonization were fixed at the parameters of the harder black body
component, representing the emission originating on or near the
surface of the neutron star. The only two remaining parameters of the
Comptonization model are the electron temperature, kT$_{\rm e}$ and optical 
depth, $\tau$ in the Comptonization region. For a given value of the 
temperature, the
upper limit on the optical depth can be computed such that the upper
limits on the hard X-ray flux (30$-$200 keV) are satisfied. Thus
temperature dependent optical depth upper limits can be obtained in the 
temperature range of interest. 

\begin{figure} 
\vspace{0.0in} 
\centerline{ 
\plotone{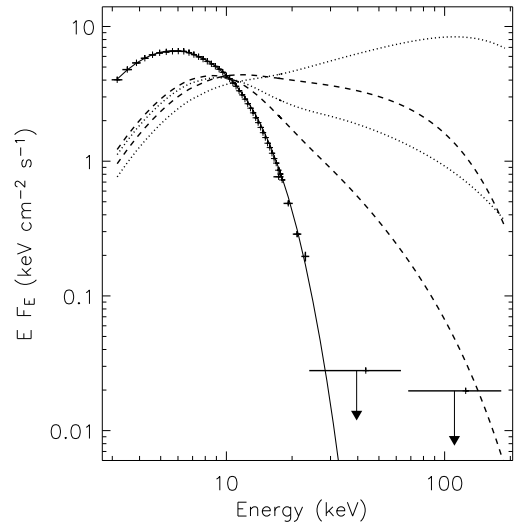}}
\vspace{0.30in} 
\caption{The broadband X-ray spectrum of GX 9+1 as seen with the RXTE/PCA
and HEXTE. The upper limits at high energies are at 2$\sigma$ level. The solid
curves are the two blackbodies that fit the PCA spectrum. The illustrative
Comptonization curves obtained with the electron cloud temperatures, 
kT$_{{\rm e}}$=30 keV (dashed) and kT$_{{\rm e}}$=60 keV (dotted), 
both using $\tau$=1 (lower curves) and $\tau$=3 (upper curves).
The higher temperature black body component, representing the neutron star 
surface emission was used as the seed for Comptonization.} 
\vspace{11pt} 
\end{figure} 

In Figure 1 we present the observed broadband X-ray spectrum of GX 9+1 
with the sum of two blackbody components (solid line) and 2$\sigma$ upper 
limits to the hard X-ray emission. Also in Figure 1, we illustrate some 
representative Comptonization spectra (obtained using COMPPS).
While performing the simultaneous fit, we fixed the electron temperature of 
the Comptonizing cloud (kT$_{{\rm e}}$) at 11 pre-selected values ranging 
from 10 to 100 keV and increasing by an increment of 10 keV, with an additional 
upper limit at kT$_{{\rm e}}$=15 keV to better understand the trend at low
electron temperatures. We determined 
the upper limits for the electron scattering optical depths ($\tau$)
of the assumed Comptonizing corona at 95\% confidence level.

\section{Results} 

We present the 2$\sigma$ upper limit values for Compton scattering optical 
depth ($\tau$) as a function of selected temperatures of scattering 
electron cloud (kT$_{\rm e}$) of GX 9+1 in Figure 2. We estimate the electron 
scattering optical depth, $\tau$ $\sim$ 0.23 even at the lowest scattering 
corona temperature considered (i.e., kT$_{\rm e}$ = 10 keV). At higher electron
temperatures the upper limit to the optical depth rapidly decreases.
This strongly indicates that for GX 9+1 the assumed corona is not optically 
thick enough to significantly alter the nature of seed photons
originating from near the stellar surface via Compton scattering.  

\begin{figure}[h] 
\vspace{0.0in} 
\centerline{\psfig{file=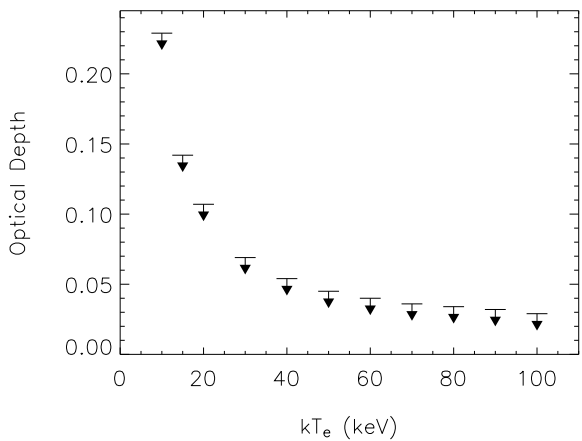,width=3.0in}}
\vspace{0.30in} 
\caption{Plot of the estimated 2$\sigma$ upper limits for the optical depth as
a function of the temperature of the electron cloud in GX 9+1.}
\end{figure} 

As an independent check, we have followed the same line of analysis
but used a different RXTE pointing observation of GX 9+1 
(Observation ID: 30042-05-02-00 performed on 27 September 1998 with an 
exposure time of $\sim$7.4 ks). We obtain a similar kT$_{\rm e}$$-$$\tau$ 
trend as seen in Figure 2.

\begin{figure}[h] 
\vspace{0.0in} 
\centerline{\psfig{file=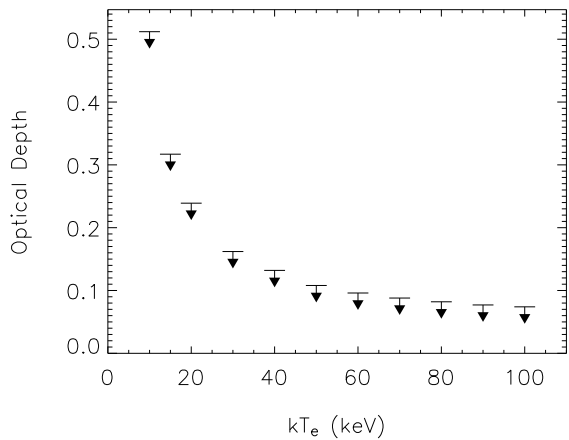,width=3.0in}}
\vspace{0.30in} 
\caption{Plot of the estimated 2$\sigma$ upper limits for the optical depth as
a function of the temperature of the electron cloud in GX 9+9.}
\end{figure} 

In Figure 3, we show the optical depth upper limits as a function of assumed 
kT$_{\rm e}$ around GX 9+9. We find that the upper limits to the 
scattering optical 
depth in this source are higher compared to that in GX9+1 at the lowest assumed
electron corona temperature. Similar to the case in GX 9+1, the upper limits 
to the optical depth fall down rapidly as the assumed electron cloud temperature
increases. We find that the optical depth of the assumed corona 
around GX 9+9 does not seem to be sufficient to suppress any beaming present
in the incoming soft X-ray radiation. 

Finally in Figure 4, we present the upper limits to $\tau$ as a function of
kT$_{\rm e}$ in Sco X-1. The overall trend is very similar to those in both
aforementioned sources. It is noteworthy here that the 2$\sigma$ upper limit
to the optical depth for the lowest electron cloud temperature of 10 keV
is almost 1.1. Such a degree of optical thickness may cause a significant
change on the properties of the incident radiation. Nevertheless, at 
electron cloud temperatures kT$_{\rm e}$ $\gtrsim$ 15 keV
the scattering optical depth upper limits drops down to about 0.6. Thus,
a possible corona around Sco X-1, if it permanently exists, is not optically
thick enough for electron scattering to efficiently play an 
important role in changing the beaming of soft X-rays emitted near the 
neutron star, unless the electron temperature is less than $\sim$10$-$15 keV.

\begin{figure}[h]
\vspace{0.0in} 
\centerline{\psfig{file=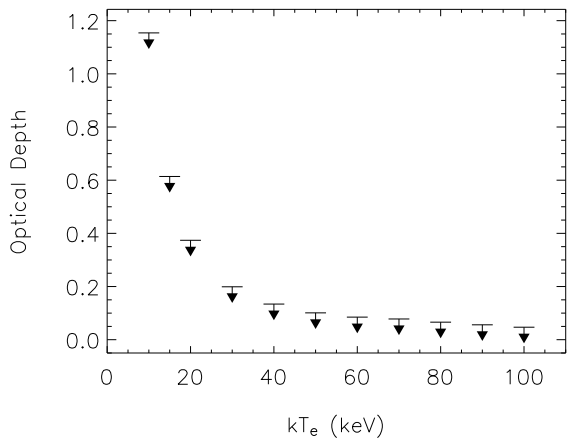,width=3.0in}}
\vspace{0.30in} 
\caption{Plot of the estimated 2$\sigma$ upper limits for the optical depth as
a function of the temperature of the electron cloud in Sco X-1.} 
\end{figure} 

As in the case of GX 9+1, we performed the optical depth estimation procedure 
using another RXTE observation of Sco X-1 as well. 
(Observation ID: 10061-01-03-00
performed on 2 January 1998 for $\sim$5.5 ks). 
We find that the 2$\sigma$ upper limit to the optical depth corresponding to
a kT$_{\rm e}$ of 10 keV as 0.96 and $\tau$ values follow a similar trend for
kT$_{\rm e}$ as was the case in Figure 4.

\section{Discussion and Conclusions} 

\subsection{Other LMXBs}

Among the LMXBs there are six Z sources, including Sco X-1, which have hard 
X-ray tails in their spectra, though only episodically 
(GX 5-1: Asai et al. 1994, Paizis et al. 2005; 
Cyg X-2: Frontera et al. 1998; Di Salvo et al. 2002; 
GX 17+2: Di Salvo et al. 2000, Farinelli et al. 2005; 
Sco X-1: D'Amico et al. 2001; GX 349+2: Di Salvo et al. 2001;  
Cir X-1: Iaria et al. 2001). In all these reported hard tail cases, the hard
X-ray spectral component is modelled with a power law (i.e., F(E) $\propto$ 
E$^{-\gamma}$ whose index ranges between -1 and 3.3. The transient appearance 
of the hard X-ray tail in these sources is not correlated with the position 
of the source radiation properties in the color-color diagram (except for
GX 17+2, Di Salvo et al. 2000, but also see Farinelli et al. 2005).

For Sco X-1, the episodic hard X-ray spectral tail was detected only in 
5 RXTE pointings and the spectra were fitted with a power law of 
indices ranging from -0.7$\pm$0.7 to 2.4$\pm$0.3 (D'Amico et al. 2001).
The 30$-$200 keV fluxes of these 5 observations range between 
8.9$\times$10$^{-10}$ and 2.1$\times$10$^{-9}$ erg cm$^{-2}$ s$^{-1}$, while
we estimate the 30$-$200 keV flux for the RXTE pointing we investigate here as
4.2$\times$10$^{-10}$ erg cm$^{-2}$ s$^{-1}$.
If there indeed exists a Compton scattering 
cloud around the neutron star and it is the cause of the episodic hard X-ray 
emission, then it may be a possible scenario for suppressing the beaming and 
pulsed signals. However, such an explanation would work only during the epochs 
with hard tails. For the rest of the time there are only upper limits
for the optical depth, which are far below the regime of efficient scattering,
indicating that the scattering corona is not likely to be thick enough to 
yield the suppression of the pulses. One is then left still needing a reason 
other than Comptonization to explain why coherent pulses are not observed.

\subsection{Comptonization Models for the Accreting Millisecond Pulsars}

There are now seven known millisecond X-ray pulsars. 
All of them have spectra 
with hard X-ray components which are fit well with Comptonization models.
For the source XTE J1751-305, Gierlinski and Poutanen (2005)
employ seed photons at the temperature $\sim $0.5 keV of the soft blackbody 
spectral component observed. 
With the COMPPS model of Comptonization they find kT$_e$ = 29$\pm$4 keV 
and $\tau$ = 1.93$\pm$0.23. If the seed photon temperature is treated 
as a free parameter, the COMPPS model gives kT$_e$ = 36$\pm$3 keV 
and $\tau$ = 1.47$\pm$0.26. The other Comptonization models tried by 
Gierlinski and Poutanen (2005) do not constrain the optical depth $\tau$  
and yield electron temperatures in the range 22-42 keV. 
For SAX 1808.4$-$3658, 
application of the COMPPS model with seed photon temperature $\sim$0.65 keV 
from the soft blackbody spectral component gives kT$_e$ = 43$\pm$9 keV and 
$\tau$ = 2.7$\pm$0.4 (Gierlinski, Done and Barret 2002). They have also
found that the Comptonized spectral component does itself pulsate. 
For XTE J1807$-$294, Falanga et al. (2005a) performed a broadband spectral
analysis using XMM-Newton, RXTE and INTEGRAL observations. They find
kT$_e$ = 37$^{+28}_{-10}$ keV and $\tau$ = 1.7$^{+0.5}_{-0.8}$ using
COMPPS with seed photon temperature of 0.75 keV. The hard X-ray spectrum of 
IGR J00291+5934 was 
adequately modelled with a different Comptonization model (COMPST, Sunyaev 
\& Titarchuk 1980) and it was found that kT$_e$ = 25$^{+21}_{-7}$ keV and 
$\tau$ = 3.6$^{+1.0}_{-1.3}$ (Shaw et al. 2005). Falanga et al. (2005b) found
significant pulsations up to $\sim$150 keV in IGR J00291+5934, indicating
phase variations of the Comptonized component. Therefore, the geometry of the 
Comptonizing region in these sources could be different than that in
non-pulsing neutron star LMXBs

Krauss \& Chakrabarty (2006) have carried out a systematic X-ray spectral
analysis of RXTE data of three millisecond X-ray pulsars and three sources 
that display no pulsations but burst oscillations. They have modeled the PCA 
and HEXTE spectra using absorbed blackbody plus Comptonization (COMPTT in 
XSPEC, Titarchuk [1994]) models. They find optical depths in the range of 
2$-$5 and plasma temperatures ranging from $\sim$20 keV to 50 keV. They find 
no distinguishing spectral properties between the coherently pulsing and
non-pulsing sources. Moreover, they also find degeneracy between the estimated
optical depths and plasma temperatures. Based on these issues, they question
the validity of the scattering scenario for the lack of pulsations.

\subsection{Conclusions} 

We revisit the Compton scattering scenario suggested earlier to
explain lack of pulsations in the majority of LMXBs. We point out that
simultaneously wiping out the pulsating signal, such a corona
would upscatter X-rays orginating from the neutron star surface
leading to appearance of a hard tail of Comptonized radiation in the
source spectrum. We utilize archival data of RXTE observations of 3
representative LMXBs (GX9+9, GX9+1 and Sco X-1) covering the full
range of atoll/Z phenomenology. Based on the upper limits on the hard
X-ray emission in the 30-200 keV energy domain and a simple but robust
method to determine the neutron star surface emission we demonstrate
that the optical depth of such a corona does not exceed $\tau\la
0.2-0.5$, unless the electron temperature is very
low, $kT_e\la 20-30$ keV. Such small values of the optical depth are
by far insufficient to suppress the pulsations.
We therefore conclude that lack of coherent pulsations can not be
attributed to the electron scatterings in the corona, at
least in the present three sources, and possibly in other LMXBs to the
extent that these sources are typical. A more likely cause of the lack
of pulsations is the absence or weakness of beaming of the X-ray
radiation emerging from the neutron star surface, caused, for example,
by the weakness of the magnetic field in high $\dot{M}$ sources. In
addition, the bending of X-rays in the gravitational field of the
compact object may play some role (Meszaros, Riffert \& Berthiaume 1988, 
{\"O}zel 2007, in preparation). Regarding the accreting millisecond X-ray 
pulsars, which exhihit pulsations in spite of evidence for Comptonization 
with $\tau$ $>$ 1, we note that the geometry of the Comptonization region must 
be different in these sources, e.g., localized near the neutron star polar 
cap, as suggested by the fact that the Comptonized spectral component in 
these sources itself pulsates.

\acknowledgments 

We thank Feryal {\"O}zel and Dimitrios Psaltis for helpful discussions, and
Rudy Wijnands for providing the database of accreting millisecond X-ray pulsars.
M.A.A. and E.G. acknowledge partial support from the Turkish Academy of 
Sciences, for E.G. through grant E.G/T{\"U}BA$-$GEB{\.I}P/2004$-$11.

\end{document}